\documentclass[aps, prl, reprint, superscriptaddress]{revtex4-1}
\usepackage{array}                                    % 提供自定义列类型
\newcolumntype{C}[1]{>{\centering\arraybackslash}p{#1}}  % 定义 C{} 列为固定宽度且居中
\usepackage{tabularx}    % 用于表格控制
\usepackage{multirow}    % 用于多行合并
\usepackage{diagbox}     % 用于斜线表头
\usepackage{makecell}    % 添加自动换行支持
\usepackage{graphicx}
\usepackage{dcolumn}
\usepackage{amsmath}
\usepackage{amssymb}
\usepackage{color}
\usepackage{bm}
\usepackage{braket}
\usepackage{txfonts}
\usepackage[colorlinks=true, letterpaper=true, pdfstartview=FitV, linkcolor=blue, citecolor=blue, urlcolor=blue]{hyperref}
\usepackage{upgreek}

\newcommand{\RNum}[1]{\uppercase\expandafter{\romannumeral #1\relax}}

\begin{document}

\title{Highly integrated broadband entropy source for quantum random number generators based on vacuum fluctuations
}

\author{Xuyang Wang}
\email{wangxuyang@sxu.edu.cn}
\affiliation{State Key Laboratory of Quantum Optics and Quantum Optics Devices, Institute of Opto-Electronics, Shanxi University, \\Taiyuan 030006, People’s Republic of China}
\affiliation{Collaborative Innovation Center of Extreme Optics, Shanxi University, Taiyuan 030006, People’s Republic of China}
\affiliation{Hefei National Laboratory, Hefei 230088, China}
\author{Yuqi Shi}
\affiliation{State Key Laboratory of Quantum Optics and Quantum Optics Devices, Institute of Opto-Electronics, Shanxi University, \\Taiyuan 030006, People’s Republic of China}
\author{Ning Wang}
\affiliation{Collaborative Innovation Center of Extreme Optics, Shanxi University, Taiyuan 030006, People’s Republic of China}
\affiliation{College of Physics and Electronic Engineering, Shanxi University, Taiyuan 030006, People’s Republic of China}
\author{Jie Yun}
\affiliation{State Key Laboratory of Quantum Optics and Quantum Optics Devices, Institute of Opto-Electronics, Shanxi University, \\Taiyuan 030006, People’s Republic of China}
\author{Jiaxu Li}
\affiliation{College of Physics and Electronic Engineering, Shanxi University, Taiyuan 030006, People’s Republic of China}
\author{Yanxiang Jia}
\affiliation{State Key Laboratory of Quantum Optics and Quantum Optics Devices, Institute of Opto-Electronics, Shanxi University, \\Taiyuan 030006, People’s Republic of China}
\author{\\Shuaishuai Liu}
\affiliation{State Key Laboratory of Quantum Optics and Quantum Optics Devices, Institute of Opto-Electronics, Shanxi University, \\Taiyuan 030006, People’s Republic of China}
\author{Zhenguo Lu}
\affiliation{State Key Laboratory of Quantum Optics and Quantum Optics Devices, Institute of Opto-Electronics, Shanxi University, \\Taiyuan 030006, People’s Republic of China}
\affiliation{Collaborative Innovation Center of Extreme Optics, Shanxi University, Taiyuan 030006, People’s Republic of China}
\author{Jun Zou}
\affiliation{ZJU-Hangzhou Global Scientific and Technological Innovation Center, Zhejiang University, Hangzhou 311215, \\People’s Republic of China}
\author{Yongmin Li}
\email{yongmin@sxu.edu.cn}
\affiliation{State Key Laboratory of Quantum Optics and Quantum Optics Devices, Institute of Opto-Electronics, Shanxi University, \\Taiyuan 030006, People’s Republic of China}
\affiliation{Collaborative Innovation Center of Extreme Optics, Shanxi University, Taiyuan 030006, People’s Republic of China}
\affiliation{Hefei National Laboratory, Hefei 230088, China}

\date{\today}

\begin{abstract}
In this work, we designed and experimentally verified a highly integrated broadband entropy source for a quantum random number generator (QRNG) based on vacuum fluctuations. The core of the entropy source is a hybrid laser-and-silicon-photonics chip, which is only 6.3 $ \times $ 2.6 $ \times $ 1.5 mm$^{3}$ in size. A balanced homodyne detector based on cascaded radio-frequency amplifiers in the entropy source achieves a 3 dB bandwidth of 2.4 GHz and a common-mode rejection ratio above 25 dB. The quantum-to-classical-noise ratio is 9.51 dB at a photoelectron current of 1 mA. The noise equivalent power and equivalent transimpedance are 8.85$\,\text{pW}/\sqrt{\text{Hz}}$ , and 22.8 k$\Omega$, respectively. After optimization using equalizer technology that eliminates the dependence of adjacent samples, the quantum random number generation rate reaches 67.9 Gbps under average conditional minimum entropy and 61.9 Gbps under the worst-case conditional minimum entropy. The developed hybrid chip enhances the integrability and speed of QRNG entropy sources based on vacuum fluctuations.
\end{abstract}

\maketitle

%%%%%%%%%%%%%%%%%%%%%%%%%%%%%%%%%%%%%%%%%%%%%%%%%%%%%%%%%%%%%%%%%%%%%%%%%%%%%%%%%%%%%%%%%%%%%%%%%%%%%%%%%%%%%%%%%%%%%%%%
%%%%%%%%%%%%%%%%%%%%%%%%%%%%%%%%%%%%%%%%%%%%%%%%%%%%%%%%%%%%%%%%%%%%%%%%%%%%%%%%%%%%%%%%%%%%%%%%%%%%%%%%%%%%%%%%%%%%%%%%
\emph{ \color{blue} 1 Introduction}

Quantum random number generators (QRNGs), which can produce true random numbers with high unpredictability, irreproducibility, and unbiasedness, are guaranteed by the basic principles of quantum physics \cite{ref1,ref2}. QRNGs play an essential role in various applications, such as quantum communication and simulations, and fundamental physical experiments \cite{ref3,ref4,ref5,ref6,ref7,ref8}. Among various ways for quantum random number generation \cite{ref9,ref10,ref11,ref12,ref13,ref14,ref15,ref16,ref17,ref18,ref19,ref20,ref21,ref22,ref23,ref24}---such as photon counting, amplified spontaneous emission, phase noise and vacuum fluctuations, and others---vacuum fluctuations offer several advantages \cite{ref16,ref17,ref18}. First, vacuum noise is a readily available source of entropy, avoiding the need for bulky external components. Second, the excess noises in a local oscillator are inherently cancelled using balanced detection, relaxing requirements on the laser and increasing the system’s resilience against external disturbances. Furthermore, the bandwidth of a balanced homodyne detector (BHD) can reach tens of gigahertz. Third, all optic devices can be integrated into chips. At present, QRNGs based on vacuum fluctuations have been extensively investigated \cite{ref25,ref26,ref27,ref28,ref29,ref30,ref31}.

The silicon photonics (SiPh) technology, which employs silicon-on-insulator (SOI) wafers as semiconductor substrate materials, is compatible with complementary metal–oxide–semiconductor (CMOS) fabrication. This means that most standard CMOS manufacturing processes can be applied, and the technology can monolithically integrate silicon electronics and photonics into the same platform \cite{ref32}. Therefore, the overall entropy source is feasible to be highly integrated.

In recent years, the SiPh based vacuum fluctuation entropy source have make significant progress, and their performances have steadily improved. In 2018, QRNG at a rate of 1.2 Gbps was realized with a 150 MHz (3 dB bandwidth) homodyne detector on a SiPh chip \cite{ref33}. In 2021, QRNG with generation rate of 18.8 Gbps based on an integrated 3.5 GHz homodyne detector was reported \cite{ref34}. In the same year, an integrated 1.5 GHz homodyne detector with improved transimpedance amplifier (TIA) circuits was designed \cite{ref35}. In 2023, the researchers achieved a quantum random number generation rate of 100 Gbps through custom codesign of optoelectronic integrated circuits and side-information reduction \cite{ref36}. In 2024, the generation rate was boosted to 240 Gbps \cite{ref37}. For SiPh integration of BHDs, 3 dB bandwidth of 1.5 GHz \cite{ref38}, 1.7 GHz \cite{ref39}, and 15.3 GHz \cite{ref40} were also reported. 

\begin{figure}[htbp]
    \centering
    \includegraphics[width=1\linewidth]{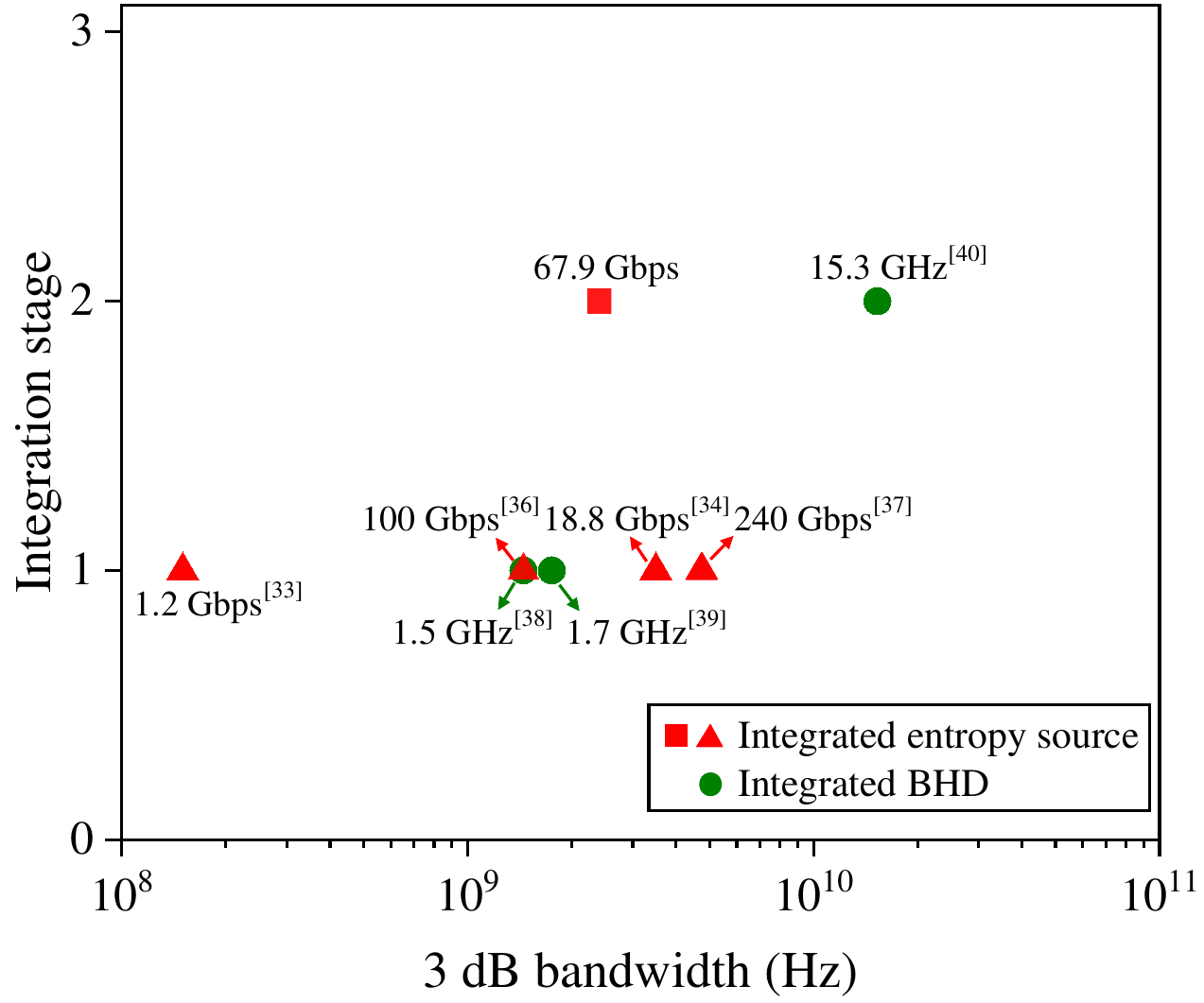}
    \caption{3 dB bandwidth of the existing integrated entropy sources and BHDs with different integration stages. The red solid square denotes the present work.}
    \label{Fig_1}
\end{figure}

Figure~\ref{Fig_1} shows the comparison of the 3 dB bandwidth (greater than 0.1 GHz) of the existing integrated entropy sources and BHDs with different integration stages. Here, we denote the integration of a 50/50 coupler, two series-connected photodiodes, and connected waveguides into a SiPh chip as “Stage 1” ; integration of any two of the three parts of the entropy sources (which are typically a laser chip, a SiPh chip, and an amplifier chip) as as “Stage 2” ; and integration of the overall entropy source into a single chip as “Stage 3” .

Notice that the entropy source of the previous high speed on-chip QRNG lacked integrated laser source, and the laser is guided into the SiPh chips through a fiber with a grating or edge coupler. Although the indirect bandgap of silicon prevents the direct generation of laser in SiPh chips, a hybrid or heterogeneous integration method can incorporate the laser and SiPh chips into a monolithicly integrated entropy source \cite{ref41,ref42}. The SiPh technology combined with hybrid or heterogeneous integration technologies can enable highrate QRNG entropy sources with small-size, low power consumption, and low cost \cite{ref26}.

In this work, we presented a highly integrated broadband QRNG entropy source based on vacuum fluctuations. The overall size of the hybrid chip, which is composed of an indium phosphide (InP) distributed feedback (DFB) laser chip and a SiPh chip, is 6.3 $ \times $ 2.6 $ \times $ 1.5 mm$^{3}$. With cascaded radio-frequency amplifiers, the BHD with a 3 dB bandwidth of 2.4 GHz and a common-mode rejection ratio (CMRR) above 25 dB was achieved. The observed quantum-to-classical-noise ratio (QCNR) is 9.51 dB at a photoelectron current of 1 mA. Based on the integrated entropy source and digital equalizer that eliminates the correlation of adjacent samples, we demonstrate an integrated QRNG with an ultrafast generation rate of 67.9 Gbps, which is the highest speed integrated QRNG at integration stage 2 (Fig.~\ref{Fig_1}).

\emph{ \color{blue} 2 Structure and packaging of the integrated entropy source}

\begin{figure}[htbp]
    \centering
    \includegraphics[width=1\linewidth]{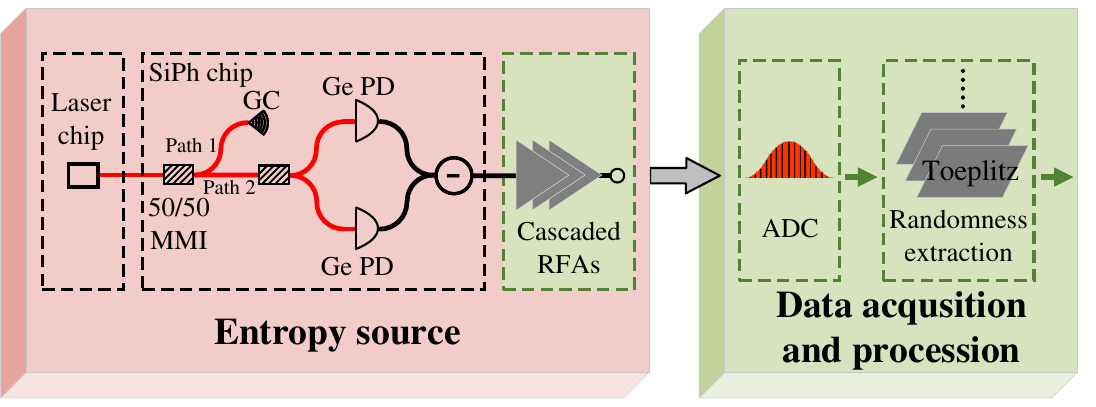}
    \caption{Structure of our QRNG based on vacuum fluctuations. GC: grating coupler, PD: photodiode, MMI: multimode interferometer, RFAs: radio-frequency amplifiers, and ADC: analog to digital converter.}
    \label{Fig_2}
\end{figure}

Figure~\ref{Fig_2} shows the structure of our QRNG based on vacuum fluctuations. The main components are the entropy source and the data acquisition and processing module. The entropy source (the core of the QRNG) comprised a laser chip, a SiPh chip, and radio-frequency amplifiers (RFAs). The entropy source outputs Gaussian white noise signals composed of shot noise (quantum noise) and electronic noise (classical noise), which are acquired by the data acquisition module through an analog to digital converter (ADC). Then, the data processing module estimates the minimum entropy and extracts the random numbers using a Toeplitz matrix.

The SiPh chip, which consists of a butt coupler, two 1 $ \times $ 2 50/50 multimode interferometer (MMI) couplers, two series-connected germanium (Ge) photodiodes, and their connected waveguides, was fabricated using the industry-standard active flow SOI technology of CUMEC. The SiPh chip is 2.5 $ \times $ 2.5 $ \times $ 0.8 mm$^{3}$ in size. The 1550 nm laser beam emitted by the InP edge-emitting DFB laser chip is coupled into the SiPh chip using the butt coupler with an insertion loss of 2.5 dB. The polarization of the laser beam is parallel to the plane of the SiPh chip and transformed into a transverse electric mode beam in the waveguide. The beam was then split by the 1 $ \times $ 2 MMI coupler. In the SiPh chip, path 1 is connected to the grating coupler for aligning in package and path 2 is used to generate shot noise. To simplify the structure, a 1 $ \times $ 2 MMI coupler with a high degree of balance is used instead of a 2 $ \times $ 2 MMI coupler, and no balance structures at the coupler outputs are required. The two output beams are directly injected into the two series-connected Ge photodiodes, each with a response of 0.9 A/W. 
\begin{figure}[htbp]
    \centering
    \includegraphics[width=1\linewidth]{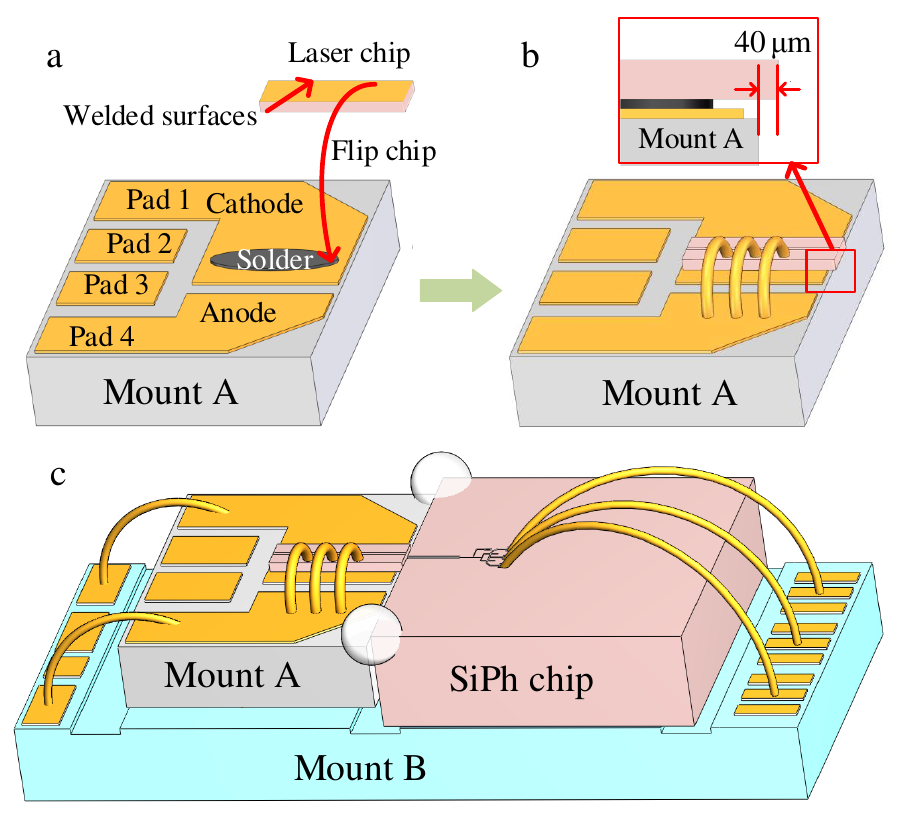}
    \caption{ Integration of the laser and SiPh chips: (a) Mount A, (b) Mount A with a flip-chipped laser chip, and (c) The packaged hybrid chip.}
    \label{Fig_3}
\end{figure}
\begin{figure*}[htbp]
		\centering
	  \includegraphics[width=1\textwidth]{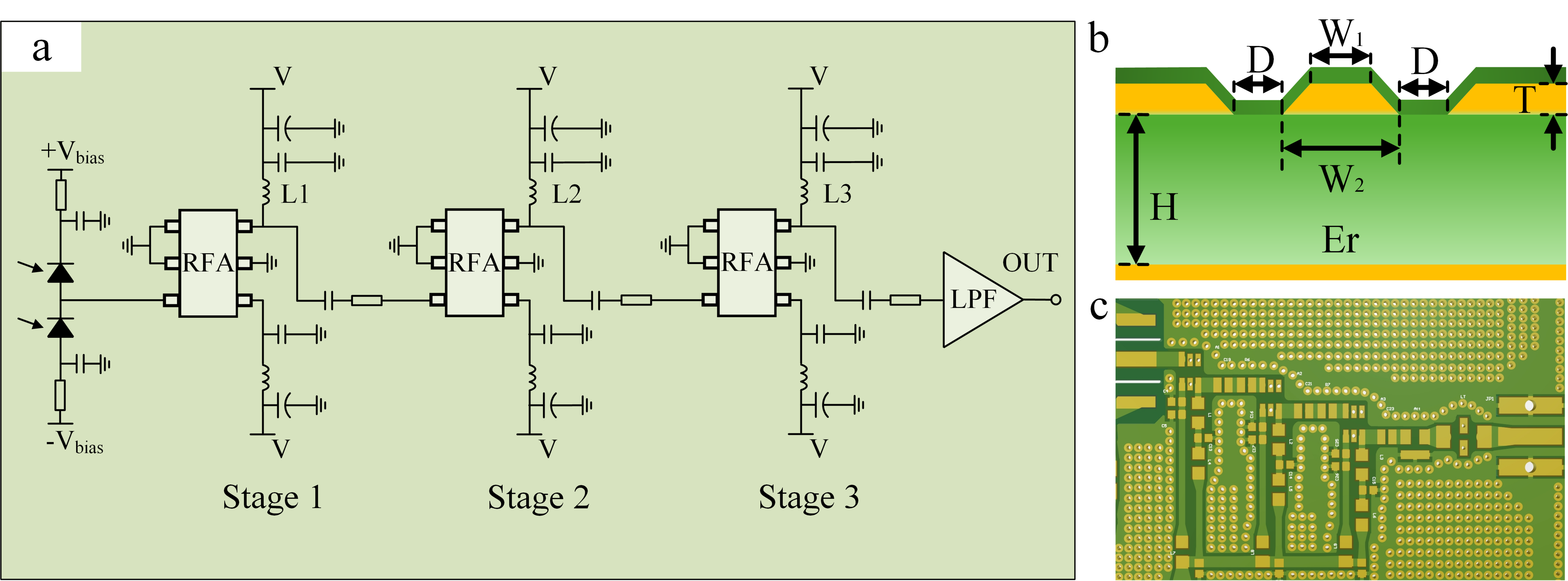}
		\caption{Circuit diagram and the layout of cascaded RFAs circuits. (a) Circuit diagram. (b) The structure of the high-frequency transmission lines. (c) The layout of RFAs circuits. LPF: Low pass filter.}
		\label{Fig_4}
\end{figure*}

To integrate the laser chip into the entropy source, we designed an aluminum nitride ceramic mount (Mount A) with good heat conduction. As shown in Fig.~\ref{Fig_3} (a). Au Pads 1 and 4 connected the DFB laser chip to the constant current sources of outside circuits, and pads 2 and 3 were used to connect a thermal resistor (which was unused here because the low heat generated by the laser chip required no temperature control). The cathode side of the laser chip was soldered to pad 1 using the flip-chip method and the anode side was connected to pad 4 via wire bonding 
(Fig.~\ref{Fig_3} (b)). The laser chip has a size of 1 $ \times $ 0.25 $ \times $ 0.12 mm$^{3}$ and protrudes 40 $\mathrm{\upmu m}$ from the edge of Mount A along the laser-emitting direction, facilitating the alignment and packaging of the hybrid chip (Fig.~\ref{Fig_3} (b), inset).

During packaging, Mount A and the SiPh chip were adhered by applying ultraviolet (UV) glue at the corners of both components, as shown in Fig.~\ref{Fig_3} (c). The UV glue seeped into the gap between Mount A and the SiPh chip, ensuring tight bonding. Heat-conducting silver glue was overlaid on the surface of Mount B, fastening the conglutination of Mount A and the SiPh chip as well as enhancing the antishaking and durability of the entropy source. After packing, the insertion loss increased to 3.1 dB. Gold-plated welding pads on Mount B connected the SiPh and laser chips to a printed circuit board (PCB).

Figure~\ref{Fig_4} presents the analog circuits of the BHD in the entropy source. The photoelectron currents difference of the two Ge photodiodes are amplified by three-stage cascaded low-noise radio frequency amplifiers (RFAs) ABA52563, fabricated with Avago’s HP25 silicon bipolar process. These monolithic silicon amplifiers are internally matched to 50 $\Omega$, and provide excellent gain with a flat broadband response from direct current (DC) up to 3.5 GHz. The cascaded RFAs are connected by capacitors, which blocked the DC voltage source V and transmits high-frequency signals, and resistors, which act as attenuators to stabilize the analog circuits. Finally, a low pass filter (LPF) with a 3 dB bandwidth of 3.5 GHz filters the high-frequency interference noises.

Obeying the PCB layout rules is essential for obtaining well-performing high-frequency amplifier circuits. The main considerations are impedance matching and parasitic capacitance and inductance. A good impedance circuit ensures the flatness of high-frequency noise signals. Herein, the sizes of transmission lines and spaces between the lines and copper coatings on the PCB board were designed using the Si9000 software. The structure and parameters of the high-frequency transmission lines are shown in Fig.~\ref{Fig_4} (b) and Tab. ~\ref{Tab.1}, respectively.

Figure~\ref{Fig_4} (c) presents the layout of cascaded RFAs circuits on the PCB. Small size electronic components were exploited to minimize parasitic inductance and optimize impedance matching. Because the sizes of many electronic devices (e.g., an LPF and a SubMiniature version A connector) did not match the width of the transmission lines, the transmission line widths were gradually varied to fit these devices while avoiding the reflection of high-frequency signals caused by drastic change. Furthermore, the space between the bias voltage pads and the common pad was hollowed to reduce the parasitic capacitance parallel to the junction capacitance of the photodiodes. The inductors L1, L2, and L3 were carefully selected as they are sensitive to parasitic capacitance. The spaces between these inductors and the copper coating were enlarged to reduce parasitic capacitance as much as possible.
\begin{table}[htbp]
    \centering
            \caption{{\small {Parameters of the high-frequency transmission lines}}}
           	    \renewcommand{\arraystretch}{1.3}
            \vspace{5pt}
		\tabcolsep 3.5pt 
			\begin{tabular}{| c | c |}
				\hline
                Substrate Height (H) & 57.66 mil\\
                \hline
				Substrate Dielectric (Er) & 4.2 \\
				\hline
				Upper Trace Width (W1) & 47.08 mil   \\
				\hline
				Lower Trace Width (W2) & 47.08 mil\\
				\hline
				Ground Strip Separation (D) & 10 mil \\
				\hline 
				Trace Thickness (T) & 1.4 mil\\
				\hline
				Impedance (Z) & 50 $\Omega$\\
				\hline
	    \end{tabular}
            \label{Tab.1}
\end{table}	

\begin{figure}[htbp]
		\centering
		\includegraphics[width=1\linewidth]{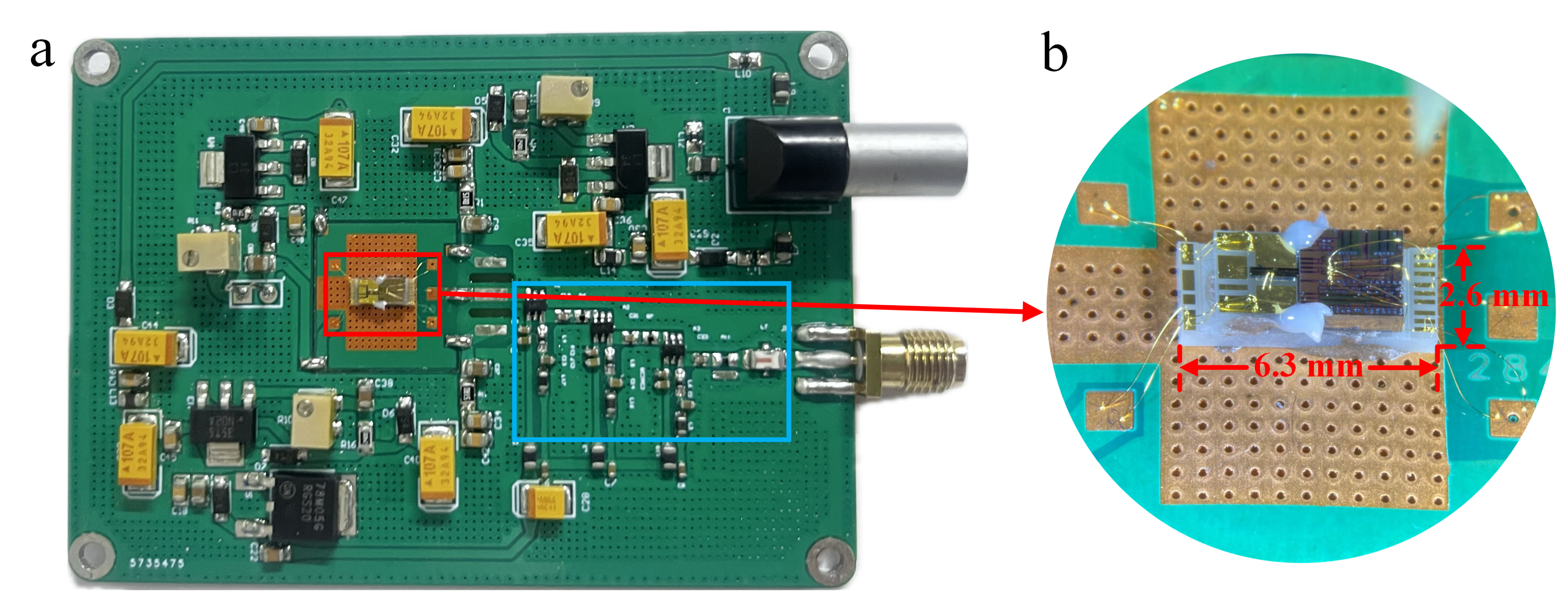}
		\caption{(a) Photograph of the integrated entropy source and its peripheral circuits and (b) microphotograph of the hybrid chip.}
		\label{Fig_5}
\end{figure}

Figure~\ref{Fig_5} (a) shows the overall integrated hybrid entropy source and its peripheral circuits. Fig.~\ref{Fig_5} (b) shows a micrograph detailing the hybrid chip composed of the laser and SiPh chips (the red rectangle in Fig.~\ref{Fig_5} (a)). The hybrid chip, with a size of 6.3 $ \times $ 2.6 $ \times $ 1.5 mm$^{3}$, was fixed on the copper surface of the PCB using silver glue to disperse the generated heat. The hybrid chip was connected to the PCB via wire bounding using 25- $\mathrm{\upmu m}$- diameter golden wires. As the laser power was low (9 mW), no temperature controller was needed. The cascaded RFAs are marked by a blue rectangle. The other circuits involved a voltage source that powered the RFAs and supplied the bias voltage to the photodiodes and a constant current source that powered the laser chip. The electronic components were soldered on the PCB using the reflow soldering technology, ensuring stable and good high-frequency performance. The whole entropy source was enclosed in a metal box to shield it from electromagnetic interference. 

\emph{ \color{blue} 3 Characterization of the integrated entropy source}

\begin{figure}[htbp]
    \centering
    \includegraphics[width=1\linewidth]{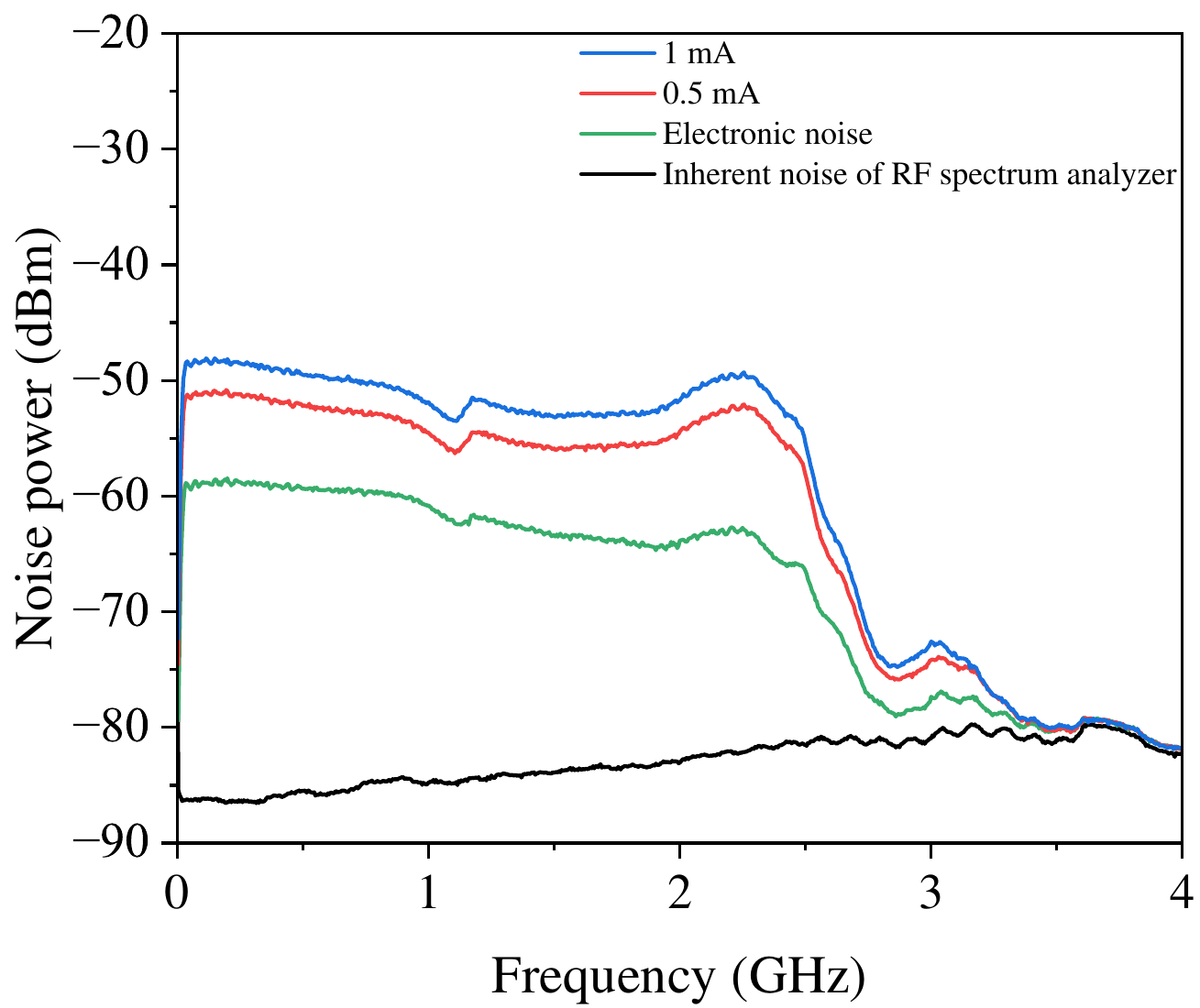}
    \caption{Noise power spectrum of the integrated BHD. The resolution bandwidth is $R_\mathrm{BW}$ = 2 MHz.}
    \label{Fig_6}
\end{figure}

The noise power sprectrum of the entropy source at different photoelectron currents from 0 Hz to 4 GHz are presented in Fig.~\ref{Fig_6}. The black line represents the inherent noise power of the RF spectrum analyzer. The green line represents the electronic (classical) noise power of the BHD, and the red and blue lines represent the noise powers at photoelectron currents of 0.5 and 1 mA. The measured shot noise was 10.5 dB above the electronic noise when the photoelectron current is 1 mA at 100 MHz. The 3 dB bandwidth of the BHD was found to be 2.4 GHz.

From Fig~\ref{Fig_6}, the noise power at 100 MHz and 1 mA was \( P_\mathrm{M} = -48.35 \, \text{dBm} \). Using the following equation:
\begin{equation}
\label{NLME}
P \, (\text{dBm}) = 10 \cdot \log_{10} \left( \frac{u^2}{R_\mathrm{Z}} \cdot {R_\mathrm{BW}}/{10^{-3}} \right),
\end{equation}
\noindent where \( P \) is the noise power at a specified frequency, \( u \) is the output-voltage noise density, and $R_\mathrm{Z}$ is the input impedance 50 $\Omega$ of the RF spectrum analyzer, the output-voltage noise density at 100 MHz was calculated to be \( u_\mathrm{M} = 6.05 \times 10^{-7} \, \text{V}/\sqrt{\text{Hz}} \). Similiarly, we can get the output-voltage noise density of the electronic noise \( u_\mathrm{E} = 1.82 \times 10^{-7} \, \text{V}/\sqrt{\text{Hz}} \) by using \( P_\mathrm{E} = -58.8 \, \text{dBm} \).

\indent The input-current noise density at photonelectron current of 1mA  is given by:
\begin{equation}
\label{LME}
i_\mathrm{Q} = \sqrt{2 \cdot q \cdot (2I_\mathrm{Q})} = 2.53 \times 10^{-11} \, \text{A}/\sqrt{\text{Hz}},
\end{equation}
\noindent where \( q \) is the charge of a single electron and \( I_\mathrm{Q} \) is the photoelectron current, the equivalent transimpedance was computed to be:
\begin{equation}
\label{ENHH}
R_\mathrm{F} = \sqrt{{u_\mathrm{M}^2 - u_\mathrm{E}^2}}/{i_\mathrm{Q}}=u_\mathrm{Q}/{i_\mathrm{Q}}=2.28 \times 10^4 \, \Omega,
\end{equation}
\noindent where $u_\mathrm{Q}$ is the noise density of shot noise.

\indent The measured noise with standard deviation $\sigma_\mathrm{M}$ at the photoelectron current of \( 1 \, \text{mA} \) comprised of shot noise with standard deviation $\sigma_\mathrm{Q}$ and electronic noise with standard deviation $\sigma_{E}$. The QCNR was then determined as:
\begin{equation}\label{tadt}
\begin{split}
\text{QCNR} 
&= 10 \cdot \log_{10} \left( \sigma_\mathrm{Q}^2 / \sigma_\mathrm{E}^2 \right) \\
&=10 \cdot \log_{10} \left( \frac{u_\mathrm{Q}^2 \cdot f_{\mathrm{3\,dB}}}{u_\mathrm{E}^2 \cdot f_{\mathrm{3\,dB}}} \right) = 9.51 \, \text{dB}.
\end{split}
\end{equation}
\indent From Eq.~(\ref{LME}, \ref{ENHH}, \ref{tadt}), it can be derived that the electronic noise is equivalent to the shot noise generated by an photoelectron current of $I_{\mathrm{E}} \approx 0.1 \, \text{mA}$. Here, a $f_{\mathrm{3\,dB}}$ = 2.4 GHz digital LPF is used to ensure that the bandwidth of electronic noise is same as the bandwidth of shot nosie. The noise equivalent power is calculated as:
\begin{equation}
\text{NEP} = i_\mathrm{E}/R(\lambda)= \sqrt{2q \cdot (2I_\mathrm{E})} / R(\lambda) = 8.85 \ \,\text{pW}/\sqrt{\text{Hz}},
\end{equation}
\noindent where \( i_\mathrm{E} \) is the current noise density of the equivalent photoelectron current \( I_\mathrm{E} \), and \( R(\lambda) \) is the responsivity of the Ge photodiode (\( 0.9 \, \text{A/W} \)) at \( 1550 \, \text{nm} \).

\indent A BHD with a high CMRR is crucial to cancel the classical intensity noise of the laser \cite{ref43}. SiPh-chip-based BHDs usually use Mach Zehnder interferometer (MZI) structures or p–i–n phase modulators to balance the two output ports of a 50/50 MMI coupler \cite{ref44}. The additional balance structures require extra balance control, which complicates the BHD structure and QRNG. In our design, an optimized symmetrical 1 $ \times $ 2 MMI structure is employed to achieve high balance degree. The balance of the two output beams of the 1 $ \times $ 2 MMI structure is less sensitive to the structure size and beam wavelength than that of the 2 $ \times $ 2 MMI structure. Furthermore, the 1 $ \times $ 2 MMI is smaller and introduces lower losses than the 2 $ \times $ 2 MMI. If a fabrication error alters the size of the 1 $ \times $ 2 MMI or the laser wavelength changes, the balance will be less affected.

\begin{figure}[htbp]
    \centering
    \includegraphics[width=1\linewidth]{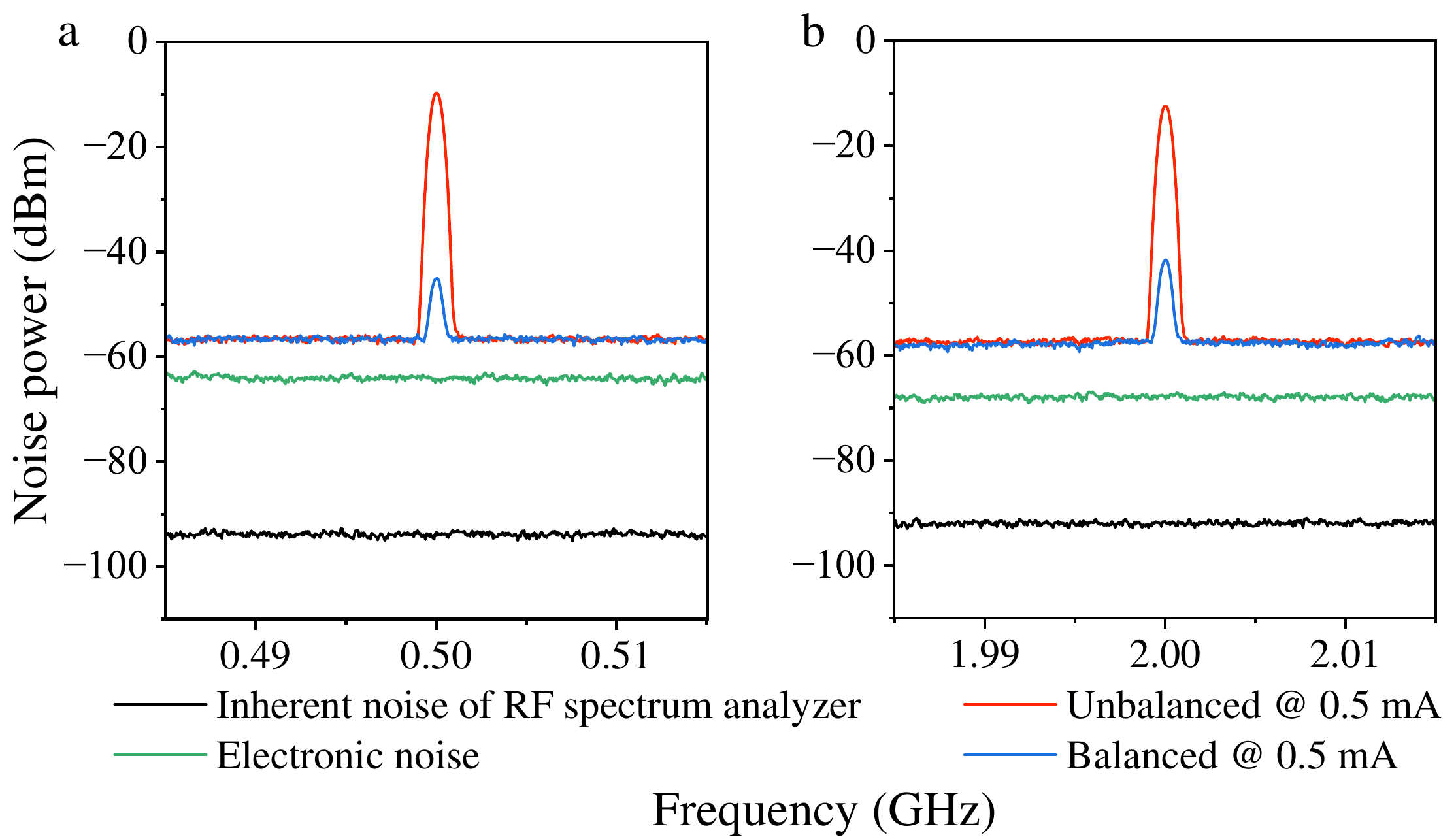}
    \caption{CMRR of the BHD measured at (a) 500 MHz and (b) 2 GHz.}
    \label{Fig_7}
\end{figure}
\indent For CMRR measurements, we directed the laser beam emitted by the DFB laser chip into an amplitude modulator and then injected the modulated beam into the integrated homodyne detector. When the photoelectron current was 0.5 mA, the CMRR exceeded 35 dB at 500 MHz (Fig.~\ref{Fig_7} (a)) and 25 dB at 2 GHz (Fig.~\ref{Fig_7} (b)).

\emph{ \color{blue} 4 Generation and analysis of quantum random numbers}

\indent Although the shot noise itself is unpredictable, it mixes with the electronic noise (classical noise) of BHD. In this case, Eve can listen or even control the electronic noise. To generate the true random number, we need to estimate the average conditional minimum entropy $\bar{H}_{\text{min}}$ and the worst-case conditional minimum entropy $H_{\text{min}}$ \cite{ref45}.
\begin{equation}
\begin{split}
\bar{H}_{\min}\left(M_{\text{dis}} \mid E\right) 
&= \lim_{\delta_{e} \to 0} \bar{H}_{\min}\left(M_{\text{dis}} \mid E_{\text{dis}}\right) \\
&= -\log_2 \Bigg[ \int_{-\infty}^{+\infty}P_E(e)\max_{m_i \in M_{\text{dis}}} P_{M_{\text{dis}} \mid E}(m_i \mid e) \, de \Bigg].
\end{split}
\end{equation}
    \begin{equation}
	H_{\min}\left(M_{\text{dis}} \mid E\right) = -\log_2 \Bigg[\max_{e \in \mathbb{R}} \max_{m_i \in M_{\text{dis}}} P_{M_{\text{dis}} \mid E}\left(m_i \mid e\right) \Bigg].
	\end{equation}
\indent Here $m$ denotes the overall output noise, and $e$ denotes the classical noise. $M_{\text{dis}}$, $E$, and $E_{\text{dis}}$ are their probability distributions. For average conditional minimum entropy, the eavesdropper can listen the classical noise with arbitrary precision, but cannot control the classical noise. In the worst-case situation, the eavesdropper not only knows the classical noise with arbitrary precision but also has the ability to control the classical noise. 
\begin{figure}[htbp]
       \centering
    \includegraphics[width=1\linewidth]{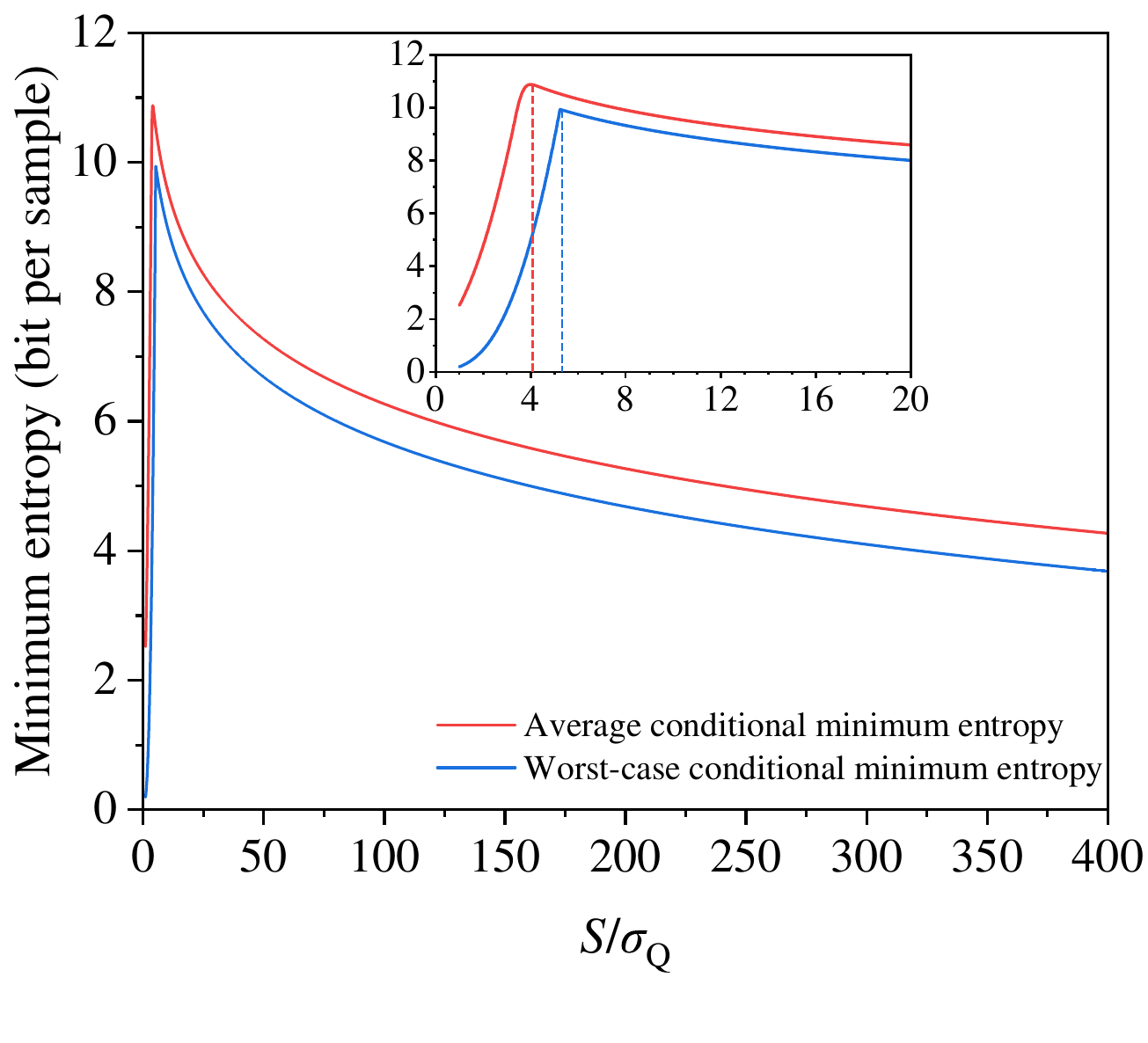}
    \caption{Average conditional minimum entropy and worst-case conditional minimum entropy vs. the ratio of ADC range to the standard deviation of quantum noise ($S/\sigma_\mathrm{Q}$). Inset shows a zoomed-in view of the peak of the minimum entropy. }
    \label{Fig_8}
\end{figure}
\begin{figure*}
		\centering
	  \includegraphics[width=1\textwidth]{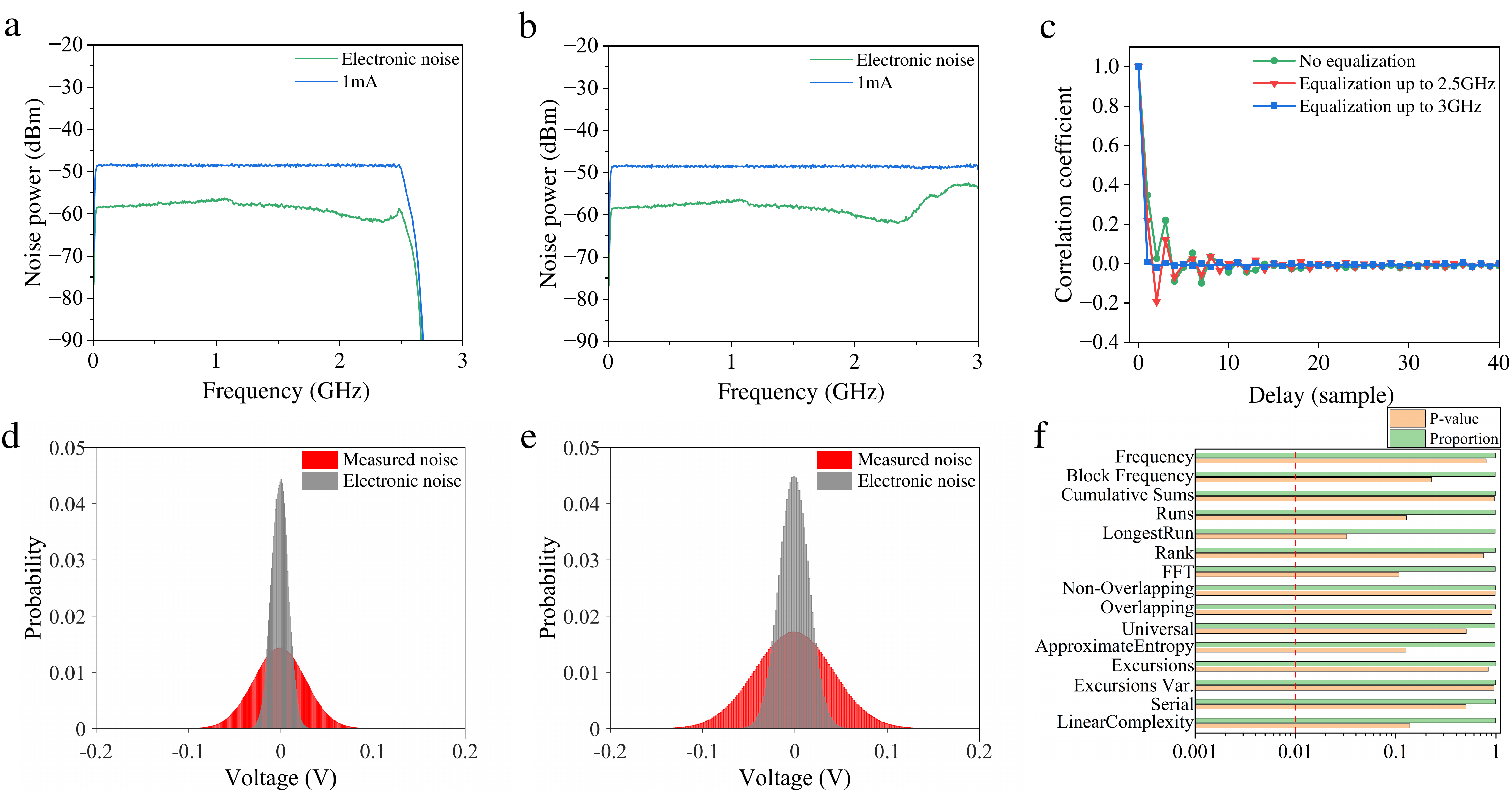}
		\caption{Equalization, correlation, and probability distribution of the data, and the randomness test results. Noise power spectra when the frequency is equalized (a) up to 2.5 GHz and (b) up to 3 GHz. (c) Effect of the equalizer on the correlation coefficients of the raw data. Histograms of measured and electronic noises before (d) and after (e) the equalization. (f) Randomness test results. }
		\label{Fig_9}
\end{figure*}

\indent The minimum entropy per sample depends on the resolution, scope, and QCNR. In our experiment, the data were acquired using an oscilloscope with a 12 bit resolution ADC. To identify the best minimum entropy, we plotted the minimum entropy as a function of $S/\sigma_\mathrm{Q}$ (Fig.~\ref{Fig_8}), where $S$ denotes the sampling scope of the oscilloscope and $\sigma_\mathrm{Q}$ = 39.3 mV is the standard deviation of quantum noise. The inset in Fig.~\ref{Fig_8} shows the peak values of $\bar{H}_{\min}$ and $H_{\min}$. As the ADC range $S$ is discrete, we selected ± $S$ = ± 160 mV ($S/\sigma_\mathrm{Q}$ = 4.07, red dashed line) to maximize the average conditional minimum entropy and ± $S$ = ± 210 mV ($S/\sigma_\mathrm{Q}$ = 5.34, blue dashed line) to maximize the worst-case conditional minimum entropy. In the calculation of minimum entropy, the QCNR is 7.34 dB which is obtained from experiment data equalized up to 3 GHz. 

\indent To maximize the data acquisition rate, we filtered the raw data acquired by the oscilloscope using equalizer \cite{ref36}. The raw data was equalized up to 2.5 GHz (Fig.~\ref{Fig_9} (a)) and 3 GHz (Fig.~\ref{Fig_9} (b)) respectively. After the equalizer, the flat white noise was obtained. The QCNR and maximum value of minimum entropy under different equalization conditions are listed in Tab.~\ref{Tab.2}. We find that the large equalization frequency range decreases the QCNR. When the equalization frequency range exceeds 3 GHz, the QCNR starts to decrease rapidly.
\begin{table}
    \centering
        \caption{{\small {QCNR, $\max(\bar{H}_{\min})$ and $\max(H_{\min})$ under different equalization conditions}}}
    \renewcommand{\arraystretch}{1.3}
    \vspace{5pt}
    \setlength{\tabcolsep}{1pt}
    \begin{tabular}{|>{\centering\arraybackslash}m{2.5cm}|>{\centering\arraybackslash}m{1.4cm}|>{\centering\arraybackslash}m{1.4cm}|>{\centering\arraybackslash}m{1.4cm}|}
        \hline
        \diagbox[width=2.58cm,height=1cm,innerleftsep=0pt,innerrightsep=0pt]{}{} & \makecell{QCNR\\ (dB)} & $\max(\bar{H}_{\min})$ & $\max(H_{\min})$ \\
        \hline
        No equalization, 2.4 GHz LPF & 9.5103 & 10.9867 & 10.0657 \\
        \hline
        Equalization up to 2.5 GHz & 9.5051 & 10.9865 & 10.0567 \\
        \hline
        Equalization up to 3 GHz & 7.3497 & 10.8770 & 9.9361 \\
        \hline
    \end{tabular}
    \label{Tab.2}
\end{table}

\indent Figure~\ref{Fig_9} (c) shows the effect of equalization on the correlations between adjacent samples. In the absence of equalization, the adjacent samples were strongly correlated. The correlation coefficients, calculated from 1.25 $ \times $ 10$^{8}$ samples, decreased when the frequency was equalized up to 2.5 GHz and nearly became zero when the frequency was equalized up to 3 GHz. The low values of correlation implies that the raw samples are close to independent and identically distributed. Panels (d) and (e) in Fig.~\ref{Fig_9} display histograms of the raw data before and after equalization, respectively. The skewness is nearly zero and the kurtosis is nearly 3 (Tab.~\ref{Tab.3}), confirming that the raw data closely follow a Gaussian distribution. $\ D_\mathrm{M}$ represents the measured noise data, and $\ D_\mathrm{E}$ represents the electronic noise data. 
 
\begin{table}[htbp]
    \centering
    \caption{{\small {Parameters estimation of kurtosis and skewness}}}
    \renewcommand{\arraystretch}{1.3}
    \vspace{5pt}
    \setlength{\tabcolsep}{2pt}
    \begin{tabular}{|>{\centering\arraybackslash}m{1.32cm}|>{\centering\arraybackslash}m{1.4cm}|>{\centering\arraybackslash}m{1.4cm}|>{\centering\arraybackslash}m{1.6cm}|>{\centering\arraybackslash}m{1.6cm}|}
        \hline
        \multirow{2}{*}{\diagbox[width=1.47cm,height=0.96cm,innerleftsep=0pt,innerrightsep=0pt]{}{}} & \multicolumn{2}{c|}{No equalization} & \multicolumn{2}{c|}{Equalization up to 3GHz} \\ 
        \cline{2-5}
        & $\ D_\mathrm{M}$ & $\ D_\mathrm{E}$ & $\ D_\mathrm{M}$ & $\ D_\mathrm{E}$ \\ 
        \hline
        Kurtosis & 2.9940 & 2.9917 & 2.9862 & 2.9863 \\ 
        \hline
        Skewness & -0.0163 & -0.0074 & -0.0090 & 0.0032 \\ 
        \hline
    \end{tabular}
        \label{Tab.3}
\end{table}

\indent Increasing the equalization frequency range further will improve the sample rate of the ADC, however, the QCNR will decrease dramatically. For the equalization of 3 GHz, our QRNG promises generation rate of 6.25 GSample/s $ \times $ 10.86 bits/sample = 67.9 Gbps (average conditional minimum entropy) and 6.25 GSample/s $ \times $ 9.9 bits/sample = 61.9 Gbps (worst-case conditional minimum entropy).

\indent By using the raw sample, the random number is extracted by constructing a Toeplitz matrix according to the estimated minimum entropy. The randomness of the extracted random number was evaluated using the NIST SP 800-22 suite \cite{ref46}. One gigabit of random numbers was extracted and tested. The randomness tests indicate good statistical characteristics of the generated random number sequences. Fig.~\ref{Fig_9} (f) shows the test results under average conditional minimum entropy.

\emph{ \color{blue} 5 Conclusions}

\indent We have designed and investigated a highly integrated broadband entropy source for QRNGs based on vacuum fluctuations. The entropy source comprises a hybrid chip and three cascaded RFAs. The hybrid chip, comprising an InP DFB laser chip (1550 nm) and a SiPh chip, is only 6.3 $ \times $ 2.6 $ \times $ 1.5 mm$^{3}$ in size. The QCNR reaches 9.51 dB at a photoelectron current of 1 mA and no temperature controller is needed. The noise equivalent power and equivalent transimpedance are 8.85$\,\text{pW}/\sqrt{\text{Hz}}$ and 22.8 k$\Omega$, respectively. Although no balancing structure was utilized, the CMRR of the BHD exceeds 25 dB. The 3 dB bandwidth of the BHD is 2.4 GHz, enabling a quantum random number generation rate of 67.9 Gbps under average conditional minimum entropy and 61.9 Gbps under worst-case conditional minimum entropy. To optimize the quantum random number generation rate, we optimize the sampling range of the ADC and utilizes the equalization technique. The randomness test results confirmed the good statistical characteristics of the generated random sequence.

\indent In future work, we will develop a hybrid chip with Stage 3 integration. In this case, the size of the entropy source can be suppressed further, combining with a higher bandwidth of the BHD, an ultrafast, compact QRNG can be achieved. We envisage that such devices will play an important role in statistical simulations, cryptography, and fundamental physical research.

\emph{ \color{blue} 6 Funding}

This work was supported in part by the Shanxi Provincial Foundation for Returned Scholars, China under Grant 2022-016, in part by the National Natural Science Foundation of China under Grant 11504219, Grant 62175138, Grant 62205188, and Grant 11904219, and in part by the Innovation Program for Quantum Science and Technology under Grant 2021ZD0300703.

%%%%%%%%%%%%%%%%%%%%%%%%%%%%%%%%%%%%%%%%%%%%%%%%%%%%%%%%%%%%%%%%%%%%%%%%%%%%%%%%%%%%%%%%%%%%%%%%%%%%%%%%%%%%%%%%%%%%%%%%
%%%%%%%%%%%%%%%%%%%%%%%%%%%%%%%%%%%%%%%%%%%%%%%%%%%%%%%%%%%%%%%%%%%%%%%%%%%%%%%%%%%%%%%%%%%%%%%%%%%%%%%%%%%%%%%%%%%%%%%%

\end{document}